\documentclass[aps,twocolumn,showpacs,floatfix]{revtex4}

\usepackage{graphicx}
\usepackage{amsmath}
\usepackage{amssymb}
\usepackage{epsf,latexsym}
\usepackage{times}
\usepackage{bm}
\usepackage[dvips]{color}

\usepackage{float}

\newcommand{\be}{\begin{equation}}
\newcommand{\ee}{\end{equation}}
\newcommand{\ba}{\begin{eqnarray}}
\newcommand{\ea}{\end{eqnarray}}

\begin{document}

\title{Knitting distributed cluster state ladders with spin chains}
\author{R. Ronke$^{1}$
}
\email{rr538@york.ac.uk}
\author{I. D'Amico$^{1}$
}
\email{irene.damico@york.ac.uk}
\author{T. P. Spiller$^{2}$
}
\email{t.p.spiller@leeds.ac.uk}

\affiliation{
$^1$ Department of Physics, University of York, York YO10 5DD, United Kingdom.\\
$^2$ School of Physics and Astronomy, E C Stoner Building, University of Leeds, Leeds, LS2 9JT.\\}
\date{\today}

\begin{abstract}
There has been much recent study on the application of spin chains to quantum state transfer and communication. Here we discuss the utilisation of spin chains (set up for perfect quantum state transfer) for the knitting of distributed cluster state structures, between spin qubits repeatedly injected and extracted at the ends of the chain. The cluster states emerge from the natural evolution of the system across different excitation number sectors. We discuss the decohering effects of errors in the injection and extraction process as well as the effects of fabrication and random errors.
\end{abstract}

\pacs{03.67.Ac, 75.10.Pq, 81.07.Vb}

\maketitle

\section{Introduction}

With conventional information processing and communications, optics provides the bandwidth and robustness for long distance communication. However, for communication over short distances, for example within and between adjacent silicon chips, signals remain electrical, to avoid the energy and cost overhead of conversion of information between different physical embodiments. Similar thinking exists in the quantum arena. Whilst quantum states of light are widely regarded as the vehicle of choice for quantum communication over large distances, there has been much recent interest in the potential use of spin chains for quantum communication over much shorter distances. When the task at hand is communication within a quantum processor, or communication between adjacent processors or registers, it may well be that a chain of spins---the same hardware from which the processors and registers are constructed---can play an effective and useful role \cite{bose2007, kay2009}.

In its simplest guise the term ``spin chain'' applies to any set of two-state quantum systems coupled to their nearest neighbours. Clearly qudits or even continuous variable oscillators could replace the qubits, but as most quantum information studies generally focus on qubits, most spin chain studies do likewise. A chain could literally comprise spins or magnetic moments, such as with a string of fullerenes \cite{twamley2003} or magnetic particles \cite{tejada2001} or nuclear spins in a molecule \cite{zhang2007}. But it could also describe a system of electrons or excitons \cite{damico2007,damico2006,niko2004} in a chain of interacting quantum dots, or other devices.

If the ground or prepared state of a spin-($1/2$) chain is all spins down ($|0\rangle$), then a complete single-qubit excitation ($|1\rangle$) is made by flipping one spin up. An arbitrary qubit state can thus be injected into a spin chain by preparing the ``injection site'' qubit in the appropriate superposition of up and down. For quantum communication, the question is as to how well this state transfers---in terms of the fidelity of the initial state against that which emerges at the ``extraction site'' at a later time, resulting from the dynamics of the chain. Usually the injection and extraction sites are the opposing ends of a chain. Study of state transfer has been made for unmodulated chains \cite{bose2003, burgarth2009_2}, systems with unequal couplings \cite{chiara2005}, systems with controlled coupling at the ends \cite{wojcik2005} and parallel chains \cite{burgarth2005}. Of specific interest to our work here is the case of linear chains where the nearest neighbour couplings $J_{i,i+1}$ between spin sites $i$ and $i+1$ are engineered to effect perfect state transfer (PST) between the injection ($i=1$) and extraction ($i=N$) sites \cite{niko2004,chris2005}. For a chain of $N$ spins the PST couplings are given by \cite{chris2005}
\begin{equation}
	J_{i,i+1}=J_{0}\sqrt{i(N-i)}\label{PST}
\end{equation}
where $J_{0}$ is a coupling that characterises the whole system and sets the timescale for PST, or mirroring, as $t_{M}=\pi \hbar/2J_{0}$.

In our work here we utilise such PST spin chains for a different purpose---the construction of distributed cluster state structures. One potential application of short range quantum communication is to build up distributed entangled resources, that can be used, or consumed, to subsequently enable distributed quantum processing through the concept of one-way computation \cite{raus2001}. In this approach, the cluster state entangled resource \cite{brie2001} is consumed by a sequence of measurements to effect the computation. Here we focus on the construction of entangled resources, which basically emerge from a suitable qubit injection and extraction protocol and the ability of PST spin chains to produce two-qubit entangling gates when operated across different excitation sectors \cite{yung2005,clark2005,clark2007}. For  such an application, there is clearly merit in being able to generate the entangled resource as rapidly and effectively as possible. We shall demonstrate that an element of a cluster state ladder can be knit in time $t_{M}$, independent of the length $N$ of the chain \cite{footnote} by suitable injection and extraction of qubits at the ends of the chain. The construction of cluster state resources using PST spin chains as an entangling bus, with assumed access to all spins in the chain, has been discussed in \cite{clark2005,clark2007}. Here we adopt the original concept of a spin chain, where access is restricted to the ends of the chain, and build the resource with a suitable injection and extraction protocol. We also consider the effects of realistic forms of decoherence acting in the spin chain and errors in the injection and extraction protocol.

\section{Spin chain dynamics}

We start with the time-independent Hamiltonian that describes the natural dynamics of a length $N$ nearest-neighbour-coupled spin chain
\begin{eqnarray}
\label{hami}
\nonumber{\cal{H}} = \sum_{i=1}^{N}E_{i}|1\rangle \langle 1|_{i} + \sum_{i=1}^{N-1} J_{i,i+1}[ |1\rangle \langle 0|_{i} \otimes |0\rangle \langle 1|_{i+1} +\\
 |0\rangle \langle 1|_{i} \otimes |1\rangle \langle 0|_{i+1}].
\end{eqnarray}
We assume that the single site excitation (from down, $|0\rangle$, to up, $|1\rangle$) energies $E_{i}$ are independent of the site $i$, or are tuned to be so. The couplings $J_{i,i+1}$ are given by (\ref{PST}). Tuning of the energies and couplings, such as via local fields or manufacturing control, is required, as per the PST scenario (\ref{PST}). The total number of excitations $T$ in the chain is given by the expectation value of the operator
\begin{equation}
\label{excitnumT}
{\cal{T}} = \sum_{i=1}^{N}|1\rangle \langle 1|_{i} .
\end{equation}
We also define the mirror operator $M$ as that which reflects the state of the chain about its midpoint (which is spin $(N+1)/2$ for odd $N$ and the ``gap'' between spins $N/2$ and $(N/2)+1$ for even $N$). So operationally $M$ effects the following to each term in any arbitrary superposition state of the chain:
\begin{equation}
\label{mirroropM}
M |a\rangle_1|b\rangle_2...|y\rangle_{N-1}|z\rangle_N=|z\rangle_1|y\rangle_2...|b\rangle_{N-1}|a\rangle_N.
\end{equation}

Clearly, for parameters restricted to achieve PST, the Hamiltonian (\ref{hami}) commutes with both ${\cal{T}}$ and $M$ and so the system energy eigenstates $|\varepsilon_k\rangle$ are also eigenstates of both ${\cal{T}}$ and $M$. For a chain of size $N$ there are $2^{N}$ eigenstates in total, with $T$ ranging from zero to $N$ and each sector $T$ containing $N!/(N-T)!T!$ eigenstates. It is also helpful to define the total chain ``spin flip'' operator ${\cal{F}}$ by:
\begin{equation}
\label{flipopF}
{\cal{F}} = \prod_{i=1}^{N} \left(|1\rangle \langle 0|_{i} + |0\rangle \langle 1|_{i}\right).
\end{equation}
The eigenstates for excitation number $N-T$ then follow from those for $T$ by application of ${\cal{F}}$. Within this framework it is straightforward to understand how PST, or more generally state mirroring \cite{albanese2004, karbach2005}, occurs. Any initial state of a spin chain $|\Psi(0)\rangle$ can be decomposed into its even and odd (under $M$) parts, so
\begin{equation}
\label{Psidecompose}
|\Psi(0)\rangle = \frac{1}{\sqrt{2}}\left(|\Psi_{+}(0)\rangle + |\Psi_{-}(0)\rangle\right)
\end{equation}
with $|\Psi_{\pm}(0)\rangle \equiv \frac{1}{\sqrt{2}}\left(|\Psi(0)\rangle \pm M|\Psi(0)\rangle\right)$. Clearly the $M$ eigenstates can be decomposed as superpositions of even and odd energy eigenstates $|\Psi_{\pm}(0)\rangle \equiv \sum_{\pm k} c_{\pm k} |\varepsilon_{\pm k}\rangle$ and then for the evolved state at time $t_M$ to have unit fidelity against the mirrored initial state $M|\Psi(0)\rangle$, it must be of the form
\begin{equation}
\label{mirrorstate}
|\Psi(t_M)\rangle =\frac{\exp(-i\theta)}{\sqrt{2}}\left(\sum_{+k} c_{+k} |\varepsilon_{+k}\rangle - \sum_{-k} c_{-k} |\varepsilon_{-k}\rangle \right).
\end{equation}
It is therefore clear \cite{albanese2004, karbach2005} that quantum state mirroring places a requirement on the chain energy level spectrum so that the phases in the evolved state conspire to give the form (\ref{mirrorstate}) at the mirror time $t_M$, with the coupling choices given in (\ref{PST}) forming an example \cite{chris2005} that produces a suitable energy level spectrum. 

The overall phase $\theta$ in the evolved state (\ref{mirrorstate}) is potentially a hindrance when it comes to PST. In addition to its time dependence, this phase depends on both the chain length $N$ and the excitation number $T$. For the case of a single excitation ($T=1$) the phase factor at the mirror time is $\exp(-i \theta(t_M)) = (-i)^{N-1}$ \cite{chris2005} and so it is recognised that $(N-1)$ needs to be a multiple of $4$ to eliminate it, although clearly in practice as $N$ will be known for a system where the couplings have been engineered to, for example, satisfy (\ref{PST}), this phase is a known correction, rather than unknown decoherence.

\section{Two-qubit gates}

When it comes to remote quantum gates, which form the basis for construction of distributed entangled resources, this phase is the enabling effect \cite{yung2005,clark2005,clark2007}. For a given spin chain of length $N$, PST does not in general work for superpositions that include states from different $T$ sectors, due to the $T$-dependence of $\theta$, although it works for arbitrary superpositions within a fixed $T$ sector. However, this effect can be turned to advantage. Consider PST spin chains where qubit states are injected onto the two extremal spins of the chain ($i=1$ and $i=N$) at time $t=0$. The initial state is thus in general a superposition of the $T=0$, $1$ and $2$ sectors. For the simplest case of the diagonal energies in (\ref{hami}) equal to zero ($E_{i}=0$) and expressed in the basis $\{|0\rangle_1 |0\rangle_N,|0\rangle_1 |1\rangle_N,|1\rangle_1 |0\rangle_N,|1\rangle_1 |1\rangle_N \}$, at time $t=t_M$ the natural evolution of the chain effects a gate $G$ given by
\begin{equation}
\label{gate}
G = \left( \begin{array}{cccc}
1 & 0 & 0 & 0 \\
0 & 0 & (-i)^{N-1} & 0 \\
0 & (-i)^{N-1} & 0 & 0 \\
0 & 0 & 0 & (-1)^N
\end{array} \right)
\end{equation}
on the initial two-qubit state. We note that for the specific case where $(N-1)$ is a multiple of $4$ the $T=1$ sector phase factors are unity \cite{chris2005}, but for all values of $N$ the gate $G$ is a maximally entangling gate, producing a concurrence of unity from an initial two-qubit product state given by an equal weight superposition of all four basis states. Thus the natural evolution of the PST spin chain enables a remote maximally-entangling two-qubit gate between the ends of the chain \cite{yung2005,clark2005,clark2007}, which cycles between four different variants with increasing $N$. Underlying all these gates is an effective phase flip that ensues in the doubly-excited sector, which can be understood as arising from the anticommutation of two non-interacting fermions as they pass through each other \cite{clark2005}. This picture is helpful in the multiple-excitation sector, where a phase factor of e$^{-i \pi}$ results from every fermionic crossing. Experimentally, this can also be achieved via crossing beams of Rydberg atoms meeting in cavities \cite{lovett2011}, which therefore represent a potential alternative medium for this procedure.

\begin{figure}
 \centering
 \includegraphics[width=0.48\textwidth]{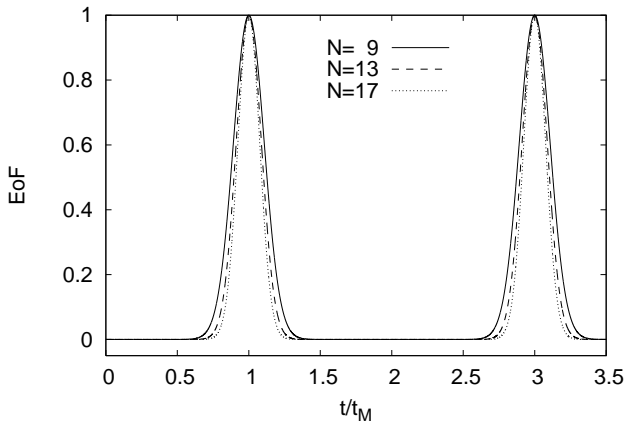}
 \caption{EoF vs. rescaled time $t/t_{M}$ for spin chains of length $N= 9, 13, 17$.}
 \label{evolution}
\end{figure}

The dynamical two-qubit gate underpins the construction of the cluster state resource, so we first examine this basic gate, acting upon qubits each prepared in the state 
$|+\rangle = \frac{1}{\sqrt{2}}(|0\rangle+|1\rangle)$ and injected simultaneously at the ends of the chain. The entangling action of the two-qubit gate is illustrated in Fig. \ref{evolution}, where the time evolution of the Entanglement of Formation (EoF) \cite{wootters1998} of the two end qubits is shown under the action of the PST spin chain dynamics. The EoF reaches unity at $t_{M}=\pi\hbar/2 J_{0}$ and every $\pi\hbar/J_{0}$ thereafter. We see that with an increasing number of spins in the chain, the width of the EoF peak decreases. However this decrease is not linear, instead the dependence of the width of peaks in Fig. \ref{evolution} with $N$ is well approximated by $1/\sqrt{N}$ within the range of chain lengths explored, demonstrating that even for very long chains recovery of the entangled qubits should be feasible.

\begin{figure}
 \centering
 \includegraphics[width=0.48\textwidth]{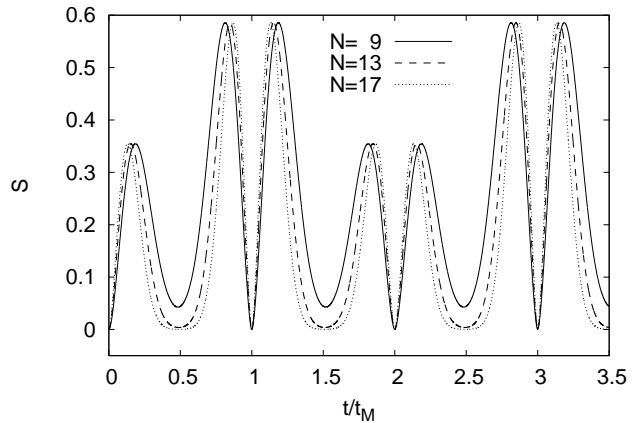}
 \caption{Entropy S vs. rescaled time $t/t_{M}$ for spin chains of length $N= 9, 13, 17$.}
 \label{fig:entropy}
\end{figure}

A single entangling gate essentially works for any length $N$ of a PST chain. However, in order to utilise additional qubit injections at the end of the chain, the length needs to be such that subsequent injections and extractions can be made independently. In order to understand this, it is helpful to consider the entropy of the end two qubits as a function of time. In order to calculate the EoF of the end two qubits, the spins comprising the rest of the chain ($i=2$ to $N-1$) are traced out, to leave the density matrix $\rho_{1,N}$ of the two end spins. If this is not pure, it provides a signature of entanglement between the two end spins and the rest of the chain. This is illustrated in Fig. \ref{fig:entropy}, which shows the evolution of the entropy 
$S=-$Tr$(\rho_{1,N} \log_{2} \rho_{1,N})$ as a function of time.

Clearly the entropy is zero at the initial injection time and all integer multiples of $t_M$. Of interest to us here is what happens in between. It can be seen from Fig. \ref{fig:entropy} that provided that $N$ is sufficiently large, the end two qubits disentangle from the rest of the chain, whilst the excitations are propagating and localised entirely in the middle region of the chain. The latter can be appreciated by looking at the site occupation probabilities, see Fig. \ref{histo}(a). This results in S $\rightarrow$ 0 for windows (that widen for increasing $N$) centred on odd half integer multiples of $t_M$. Similar to Fig. \ref{evolution}, the width of the dip at $t=t_M$ decreases approximately as $1/\sqrt{N}$ within the range of chain lengths explored. For chains of length $N \geq 9$, it is possible to independently inject further qubits. This will form the basis of the cluster ladder construction. Before considering the injection and extraction protocol, though, we first consider the effects of various errors on the basic entangling gate and the evolving two-end-qubit entropy.

\section{Errors and decoherence}

\begin{figure}
 \centering
 \includegraphics[width=0.48\textwidth]{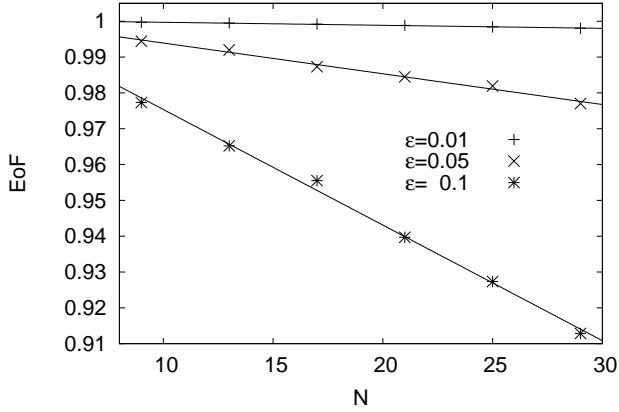}
 \caption{EoF at $t=t_{M}$ vs. $N$ for three values of $\varepsilon$, as labelled. Data points are averaged over 100 random on-site energy realisations.}
 \label{linear}
\end{figure}

For non-zero but site-independent excitation energies for the qubits in the chain (the diagonal terms in Eq. (\ref{hami})), additional phases arise in the chain dynamics, but these do not affect the entangling capacity of the remote two-qubit gate. Of more interest is the effect of variations, or errors in the site energies. A simple estimate of the effect of errors in the energy level spectrum \cite{chris2005} suggests an overall error, or loss in fidelity for PST that scales linearly with $N$. Figure \ref{linear} illustrates the EoF evolution as a function of the spin chain length $N$, including a random modulation of the on-site energies $E_i$. The on-site energies are now given by $E_i=\varepsilon r_i$, 
where $0\leq r_i \leq  1$ is a random number from a uniform distribution. Each point in  Fig. \ref{linear} corresponds to the EoF at $t=t_M$ averaged over 100 random realizations.
 Figure \ref{linear} confirms that the loss in EoF scales linearly with the number of spins in the chain. Moreover, it shows that the EoF decay is not linear for increasing $\varepsilon/J_{max}$, with $J_{max}=max_i\{J_{i,i+1}\}=1$. For small and medium perturbations of the on-site energies, EoF close to unity can still be achieved for all chain lengths considered.

As we are dealing with multiple excitations in the chain, we also consider the effect of interaction between excitations in nearby sites by adding the following term to Eq. (\ref{hami}): 

\begin{equation}
 \label{interc}
{\cal{H}}' = \sum_{i=1}^{N-1} \gamma J_{0} |1\rangle \langle1|_{i} \otimes |1\rangle \langle1|_{i+1}.
\end{equation} 

\begin{figure}
 \centering
 \includegraphics[width=0.48\textwidth]{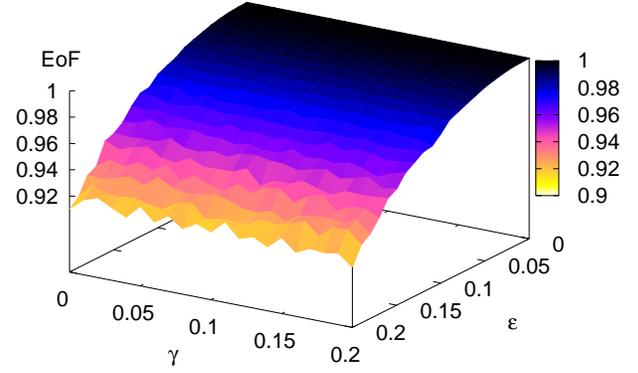}
 \caption{EoF at time $t=t_{M}$ for a 9-spin chain vs $\gamma$ and $\varepsilon$. Data points are averaged over 100 random on-site energy realisations.}
 \label{decay}
\end{figure}

For example, ${\cal{H}}'$ may correspond to a biexcitonic interaction in quantum dot-based chains \cite{damico2001, rinaldis2002}. Figure \ref{decay} shows the dependence of the EoF at $t=t_M$ on the magnitude of the maximum on-site energy perturbation $\varepsilon$ and of the site-interaction perturbation $\gamma$ for $N=9$. Even with perturbations as big as 20\% of the characteristic coupling strength $J_{0}$, the EoF hardly suffers and remains above 90\%, indicating that the system is extremely robust against these sorts of errors. The influence of $\gamma$ in Fig. \ref{decay} is limited, as three quarters of the initial input state contain no more than one excitation and are thus not subject to the effects of interactions between excitations.

\begin{figure}
 \centering
 \includegraphics[width=0.48\textwidth]{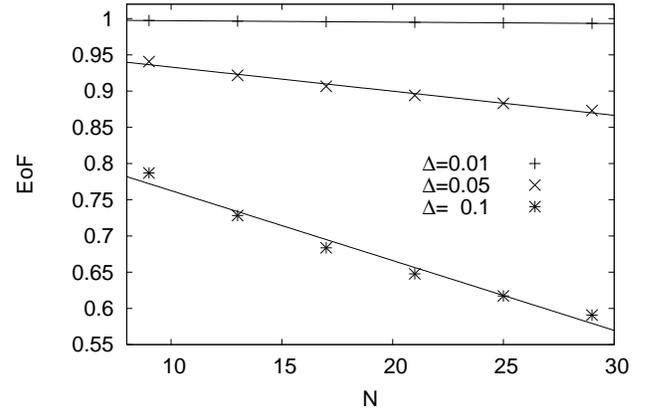}
 \caption{EoF at $t=t_{M}$ vs. $N$ for three values of $\Delta$, as labelled. Lines are best fit to numerical data.}
 \label{fig:nnni2}
\end{figure}

Next we explore the effect on the remote gate of unwanted longer range interactions \cite{avellino,kay2006}, which could be an issue when considering pseudospins based on charge degrees of freedom. We add to Eq. (\ref{hami}) the perturbative term
\begin{eqnarray}
\label{hami''}
\nonumber{\cal{H}}''= \sum_{i=1}^{N-2} J_{i,i+2}[ |1\rangle \langle 0|_{i} \otimes |0\rangle \langle 1|_{i+2} +\\
 |0\rangle \langle 1|_{i} \otimes |1\rangle \langle 0|_{i+2}],
\end{eqnarray}
with $J_{i,i+2}=\Delta(J_{i,i+1}+J_{i+1,i+2})/2$ to simulate the original coupling modulation. A discussion about the experimental relevance of this term has been given in \cite{me1} on the example of graphene and self-assembled quantum dots. ${\cal{H}}''$ commutes with both ${\cal{T}}$ and $M$, so that a set of common even and odd eigenstates still exists.
The dependence of the EoF at $t=t_M$ as a function of $N$ is presented in Fig. \ref{fig:nnni2}, for three different values of $\Delta$. The EoF displays a decay linear in $N$. For small values of $\Delta$, relevant e.g. to graphene quantum dots \cite{me1}, the EoF is well conserved, although the effect is much more pronounced in very long chains as the increase in number of spins leads to more perturbation terms in Eq. (\ref{hami''}). For example, with a value of $\Delta$ as large as 0.1, a short chain ($N=9$) achieves an EoF of 0.79, with this dropping to 0.59 for a long chain ($N=30$), as shown in Fig. \ref{fig:nnni2}. We also note that for $\Delta=0.1$, while PST can be qualitatively measured at $t=t_M$, its subsequent periodicity is lost (not shown). This implies that in this case the gate should be put into effect at $t=t_M$.

Another potential source of gate error for the remote two-qubit gate is a timing error in the ``extraction'' of the qubits that have undergone the gate. Clearly from the parabolic expansion of the concurrence around its maximum at $t=t_M$, this error is second order in any timing error $\delta t$, consistent with the PST timing error \cite{chris2005}.

\section{Distributed cluster ladder construction}

\begin{figure}
 \centering
 \includegraphics[width=0.48\textwidth]{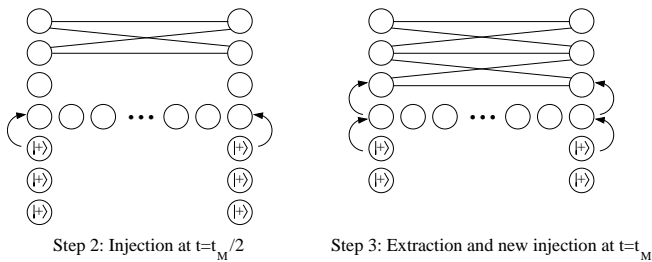}
 \caption{Diagram of steps 2 and 3 of the cluster ladder knitting protocol. Solid lines indicate edges of the resulting cluster state}
 \label{fig:knitting}
\end{figure}

Having discussed the building block gate in detail, we now turn our attention to the repeated use of this operation to generate a cluster ladder. A cluster state resource \cite{brie2001} between multiple qubits is realised by placing all qubits in state $|+\rangle$ and then applying a controlled-phase entangling gate between all pairs of qubits that are to be connected in the layout of the cluster state. The following protocol of injection and extraction for a PST spin chain generates a cluster ladder:
\begin{enumerate}
\item $t=0$: Inject $|+\rangle$ at each end qubit ($i=1$ and $i=N$) to a ground state chain (all zeros). This could be performed by a SWAP operation with register qubits adjacent to each end site prepared in state $|+\rangle$. (Clearly the SWAP operation has to be fast on the timescale set by $t_M$.)
\item $t=t_M/2$: Inject $|+\rangle$ at each end qubit, which at this time will be disentangled from the rest of the chain (see Fig. \ref{fig:entropy}).
\item $t=t_M$: Extract the end qubits and inject $|+\rangle$ at each end qubit. 
\item Repeat last step as many times as desired, at time intervals of $t_M/2$.
\item Extract the last two qubits in the ladder when they reach the ends of the chain.
\end{enumerate}
It is assumed that each time a pair of qubits in the cluster ladder is extracted, each qubit is shuttled along a register at each end, with the next pair of $|+\rangle$ states moved into positions ready to be swapped in. This is illustrated in Fig. \ref{fig:knitting}.

Step 2 of the protocol, the second injection of $|+\rangle$ states, relies on the fact that the end qubits are initially empty. As especially for short chains, this cannot be guaranteed, we would then have some unwanted entanglement between the end qubits and the auxiliary qubits which carried the $|+\rangle$ states. If we assume that the injection is done by a SWAP operation, we can however refocus the system by measuring the auxiliary qubits: finding no excitations corresponds to a successful injection and also collapses the unwanted entanglement, whereas finding two excitations in the auxiliary qubits means that we have extracted all the excitations which the chain contained previously and are now back to the state of the system as after step 1 of the protocol, allowing us to proceed and attempt another set of injections a time $t_M /2$ later. In the event that we measure a single excitation in only one of the auxiliary qubits, the system cannot easily be recovered to a form which would allow us to continue with our protocol and needs to be re-initialised.

\begin{figure}
 \centering
 \includegraphics[width=0.48\textwidth]{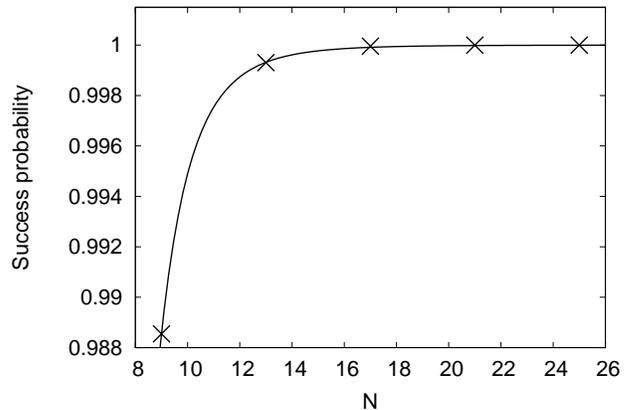}
 \caption{Probability of successfully injecting at $t=t_M/2$ vs. chain length $N$. Data fitted as a function in $1/N$.}
 \label{fig:prob}
\end{figure}

The probability of successfully injecting a pair of $|+\rangle$ states at $t_M /2$ is however extremely high: Fig. \ref{fig:prob} shows the probability of a successful injection with increasing $N$ and we see that even for $N=9$, the success probability is already next to 99\%, with the failure probability (the deviation from unity) further decreasing as $1/N$. These results allow us to be very confident about the efficiency of our proposed protocol.

It must be noted that other injection methods, such as tunnelling for example, might not allow for a refocussing mechanism as described above. Injection methods which can result in one or both $|+\rangle$ states remaining in the auxiliary qubits will lead to retention of the unwanted entanglement between the chain and the auxiliary qubits as measurement of the auxiliary qubits (which, in the case of a failed injection, are in a $|+\rangle$ state) does not lead to conclusive results. In this work we will therefore only consider injection of the SWAP type with refocussing taking place immediately after injection.

An equivalent refocussing for the potentially imperfect extraction of $|+\rangle$ states at $t_M$ (step 3 of the protocol) is however not possible. If the generated $|+\rangle$ states are extracted into empty storage qubits via a SWAP operation, while there is no unwanted entanglement between the chain and the storage qubits, there is no measurement we can perform to check whether the extracted states are indeed $|+\rangle$ states without destroying them. This might lead to some inaccuracies for very long cluster state ladders but results below suggest that this error does not limit practical use of the protocol. 

\begin{figure}
 \centering
 \includegraphics[width=0.48\textwidth]{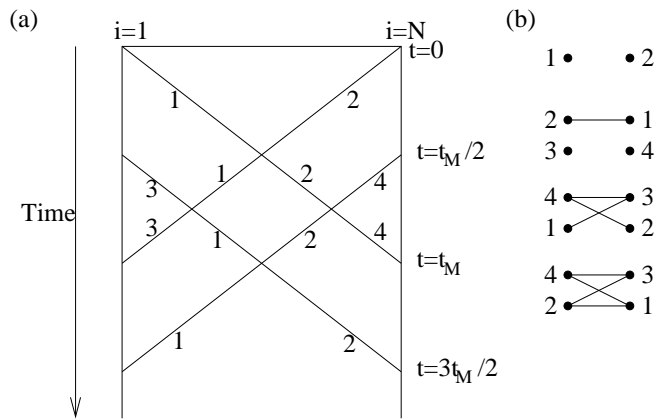}
 \caption{(a) Propagation of the four qubits involved in the generation of a crossed square cluster state. (b) Schematic formation of the crossed square cluster state.}
 \label{fig:squarecluster}
\end{figure}

In order to now demonstrate our protocol, we consider the generation of a `crossed' square cluster state, which simply involves injection of two pairs of qubits at $t=0$ and $t=t_{M}/2$ with subsequent refocussing and their extraction at $t=t_{M}$ and $t=3 t_{M}/2$ (see Fig. \ref{fig:squarecluster}). First of all, we are dealing with a chain of length $N$ such that $(N-1)$ is a multiple of 4, so the gate in action in this case is, from Eq. (\ref{gate}),
\begin{equation}
\label{gate2}
G' = \left( \begin{array}{cccc}
1 & 0 & 0 & 0 \\
0 & 0 & 1 & 0 \\
0 & 1 & 0 & 0 \\
0 & 0 & 0 & -1
\end{array} \right).
\end{equation}
It is worth noting that $G'$ is the product of a CZ gate, which leads to the desired entanglement, and a SWAP gate. In Fig. \ref{fig:squarecluster} (a), this is illustrated by the labels on the lines representing excitations swapping as they cross whereas Fig. \ref{fig:squarecluster} (b) shows how the entanglement between the individual qubits it built up. Let us now look at the generation of a crossed square cluster state that this figure illustrates.

\begin{figure}
 \centering
 \includegraphics[width=0.48\textwidth]{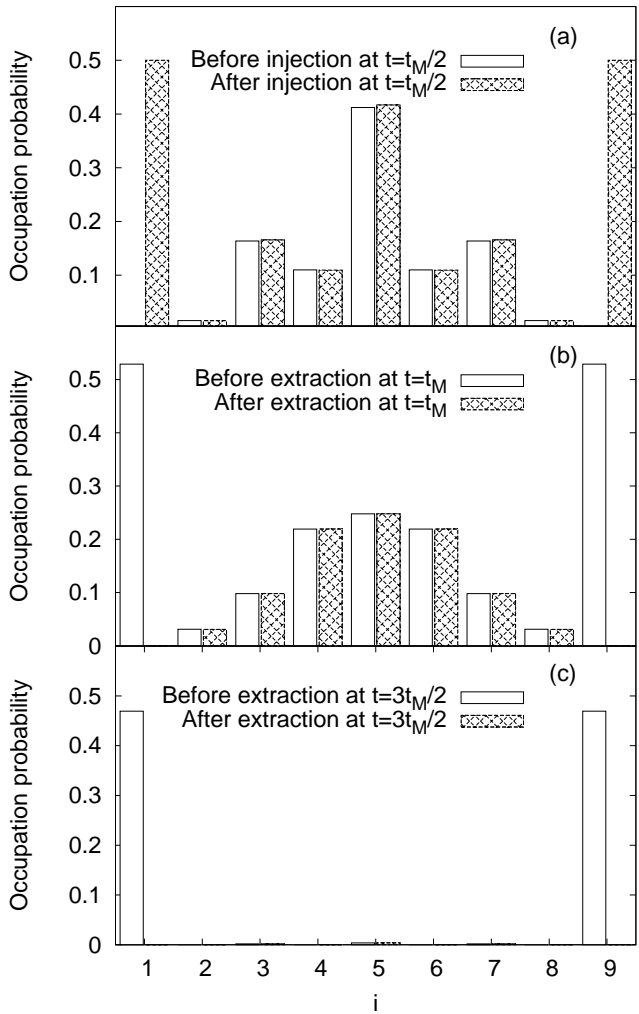}
 \caption{Changes in occupation probability of site $i$ for a 9-spin chain during the generation of a crossed square cluster state.}
 \label{histo}
\end{figure}

To construct a crossed square cluster state, we inject excitations 1 and 2 at sites $i=1$ and $i=N$ respectively and then wait for $t_{M}/2$ until we know that these sites are (nearly) completely disentangled from the rest of the chain. We then inject excitations 3 and 4 at sites $i=1$ and $i=N$ respectively and refocus. The change in occupation probability of the individual spins can be seen in Fig. \ref{histo}(a) on the example of a 9-spin chain, for which the refocussing success probability is 0.9885.

The occupation probability is defined as follows: let us assume that the set $\{|\phi_i\rangle\}$ of $k$ basis vectors forms the basis for our spin chain, such that any spin chain state $|\psi\rangle$ can be written as $|\psi\rangle = \sum^{k}_{i=1}c_i |\phi_i\rangle$, with $\sum^{k}_{i=1}|c_i|^2=1$. The basis vectors can be represented as $|\phi_i\rangle = |j_1 j_2 \cdots j_N\rangle$, with $j=\{0,1\}$. Those basis vectors which contribute to the occupation of a site $s$ are then $|\phi_{i,s}\rangle = |j_1 j_2 \cdots 1_s \cdots j_N\rangle$, where $1\leq s \leq N$. Each $|\phi_{i,s}\rangle$ is weighed by its coefficient $c_{i,s}$ and so the occupation probability of a site $s$ is given by $\sum_{i=1}^{k} |c_{i,s}|^2$. As a result, it is to be expected that the total area of the histogram representing the occupation probability is equal to the number of excitations in the system, so $\sum^{N}_{s=1} |c_{i,s}|^2 = T$ for a specific excitation sector.

In Fig. \ref{histo}(a) we see how inaccuracies in manipulation of the spin chain lead to deviations from the ideal scenario. As the entropy of the depicted 9-spin chain does not reach zero at $t_M/2$ (see Fig. \ref{fig:entropy}), the extremal spins $1$ and $N$ are not entirely decoupled from the rest of the chain. Despite refocussing after injection, which guarantees an occupation probability of exactly $0.5$ of spins $1$ and $N$ after injection at $t_M/2$ (frame (a)), we see in frame (b) that the occupation probability of spins $1$ and $N$ before extraction at $t=t_M$ is bigger than $0.5$. Consequently, in frame (c) the occupation probability of spins $1$ and $N$ before extraction at $3t_M/2$ is smaller than $0.5$. As we will see below, this will lead to a small loss in quality of the produced crossed square cluster state. 

After $t=t_M/2$, we have injected all the necessary excitations and wait for another $t_{M}/2$ until $t=t_{M}$ before using a SWAP operation to extract excitations 3 and 4 at their injection sites, where the swapping qubits which are not part of the chain are presumed initially empty and now serve as storage. At this stage, the four excitations are already entangled, as can be seen in Fig. \ref{fig:squarecluster} (b). Finally, at $3t_{M}/2$ we extract excitations 1 and 2 at their initial injection sites via another SWAP operation, thus completing the cluster state. Again, the auxiliary qubits used to operate this SWAP are presumed initially unexcited and now serve as storage. The dynamics of these qubits is included in the simulation. Figure \ref{histo}(c) illustrates that following this operation, the chain is virtually empty and thus ready for use again. We note that the residual occupation in the chain is related to the imperfect disentanglement for a 9-spin chain of the end qubits at $t=t_{M}/2+nt_{M}$, with $n=0,1,2,\cdots$. The end qubits disentangle virtually perfectly as longer chains are considered, see Fig. \ref{fig:entropy}.

\begin{figure}
 \centering
 \includegraphics[width=0.48\textwidth]{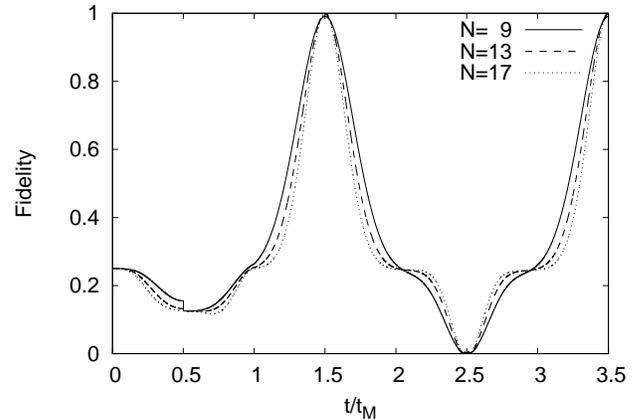}
 \caption{Fidelity of the ideal state vs. rescaled time $t/t_{M}$ for spin chains of length $N= 9, 13, 17$ used to build a crossed square cluster state.}
 \label{fid_2}
\end{figure}

Monitoring the quality of even just this four-qubit cluster state cannot be achieved via the EoF, which is only suitable as a measure of bipartite entanglement, but we instead consider the fidelity $F$ against the ideal cluster state $\psi_{ideal}$ that we are hoping to achieve:
\begin{equation}
	F=|\langle \psi_{ideal} |e^{-i{\cal{H}}t/\hbar}| \psi_{ini} \rangle|^{2},
\end{equation}
so perfect cluster state construction is achieved when $F=1$. As in fact we are more interested in the true nature of the achieved state rather than simply in the amount of entanglement produced, this measure is very suitable for our purposes. Figure \ref{fid_2} shows the evolution of this fidelity during and after implementation of the construction protocol for a crossed square cluster state, without the final read-out step at $t=3t_{M}/2$ so that the evolution of the achieved cluster state continues periodically beyond this time, achieving the same maximum fidelity at every $2t_M$ after $t=3t_{M}/2$. Notice the step at $t=t_{M}/2$, which corresponds to the injection of the second pair of excitations. This slight discontinuity is most clearly visible for $N=9$, where we know from Fig. \ref{fig:entropy} that the dip in entropy is not as low as for longer chains. Nonetheless, $N=9$ achieves a fidelity of 0.9915 at $t=3t_{M/2}$ and the two longer chains pictured both achieve unity at this time. This confirms that a 9-spin chain is long enough to demonstrate the effects we wish to highlight and so we will use this chain as our default device when considering a single value of $N$ only. Refocussing for $N=13$ and $N=17$ is done with success probabilities of 0.9993 and 1.000 respectively, showing that modest increases in $N$ lead to both higher fidelity and higher refocussing success probability. As in Fig. \ref{fig:entropy}, if no excitations are read out at $t=3t_{M/2}$, the evolution in Fig. \ref{fid_2} continues to be periodic over $2t_{M}$.

\begin{figure}
 \centering
 \includegraphics[width=0.48\textwidth]{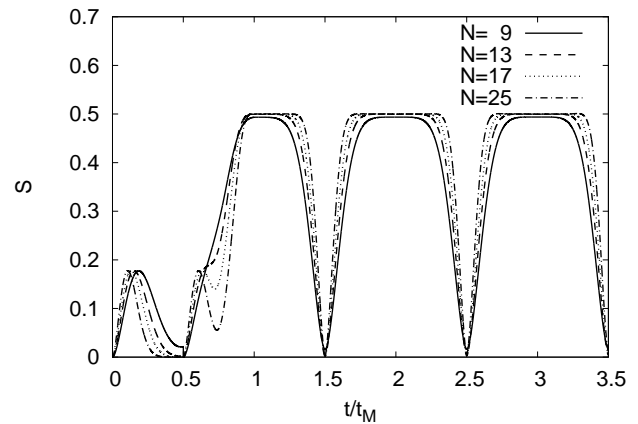}
 \caption{Entropy S vs. rescaled time $t/t_{M}$ for spin chains of length $N= 9, 13, 17, 25$ used to build a crossed square cluster state.}
 \label{VNE_2}
\end{figure}

We can confirm this result by also monitoring the entropy S, with the difference that instead of using the density matrix $\rho_{1,N}$ of the two end spins, we use the density matrix of the four storage qubits, i.e. the two end qubits plus the auxiliary qubits. This is illustrated in Fig. \ref{VNE_2}, which clearly shows the entropy S dipping to zero (or virtually zero for $N=9$). Again, the final read-out step at $t=3t_{M}/2$ is omitted, leading to a continuous evolution of the constructed cluster state which displays dips in entropy every $t_M$ after the initial dip at $t=3t_{M}/2$. Analogous to the dip width decrease observed in Fig. \ref{fig:entropy}, the variation in width of the dip at $1.5t_M$ is approximated by $1/\sqrt{N}$.

\begin{figure}
 \centering
 \includegraphics[width=0.48\textwidth]{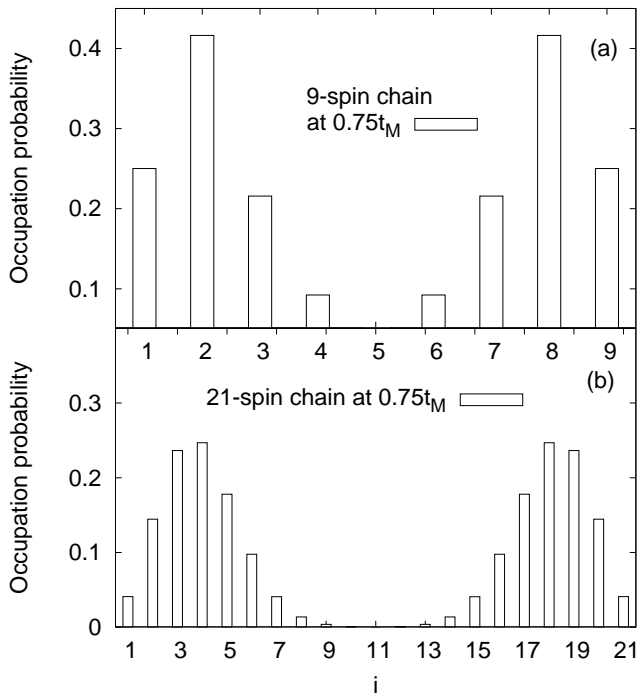}
 \caption{Occupation probabilities of site $i=1, \cdots N$ at $t=0.75t_M$ for a 9-spin chain (a) and a 21-spin chain (b).}
 \label{histo_dip}
\end{figure}

S displays an additional dip at $t=0.75t_M$, which is almost unnoticeable for short chains such as $N=9$ but tends to zero for very long chains such as $N=25$. This corresponds to a temporary decoupling of the end spins (sites 1 and $N$) as the excitations pass through each other (as indicated by the intersecting lines in Fig. \ref{fig:squarecluster}). To illustrate this, we record the site occupation probabilities at $t=0.75t_M$  in Fig. \ref{histo_dip} for a short chain ($N=9$ in panel (a)) and a long chain ($N=21$ in panel (b)). We see that for $N=9$, where we see no dip at $t=0.75t_M$ in Fig. \ref{VNE_2}, there is a high occupation probability for the ends spins, whereas for $N=21$, where we have a significant dip in entropy S at $t=0.75t_M$, the occupation probability of the end spins is very low, leaving the spins nearly decoupled from the rest of the chain. We also note that regardless of the chain length, the middle of the chain is empty at this time.

\section{Errors in the cluster knitting protocol}

Just like the two qubit gate in section II, the cluster knitting protocol we presented will be subject to various unpredictable errors and decoherences. Due to its higher complexity, the knitted cluster state is potentially more affected by defects similar to the ones discussed in section IV. Due to computational restrictions, the more involved nature of the numerical simulations presented in this section also puts a limit on the lengths of spin chains we can investigate. The longest spin chain we consider here will therefore have 25 spins and not 29 as in section IV.

\begin{figure}
 \centering
 \includegraphics[width=0.48\textwidth]{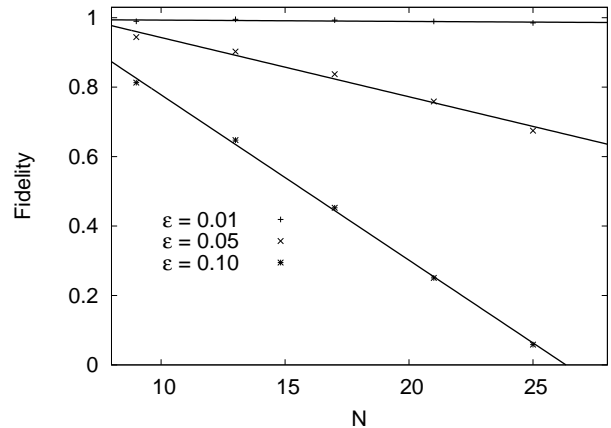}
 \caption{Fidelity of the ideal state at $t=3t_{M}/2$ vs. $N$ for three values of $\varepsilon$, as labelled, where the chains are used to build a crossed square cluster state. Data points are averaged over 100 random on-site energy realisations.}
 \label{enercN_2}
\end{figure}

\begin{figure}
 \centering
 \includegraphics[width=0.48\textwidth]{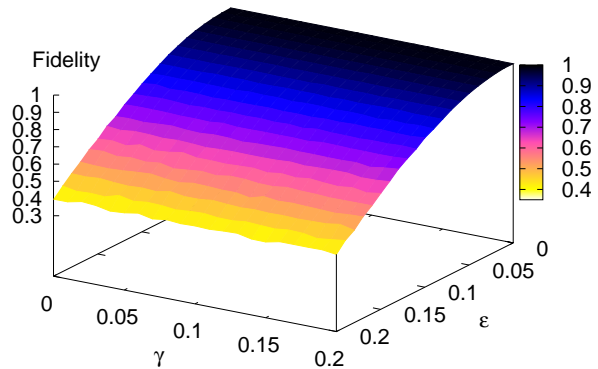}
 \caption{Fidelity of the ideal state at time $t=3t_{M}/2$ for a 9-spin chain used to build a crossed square cluster state vs $\gamma$ and $\varepsilon$ .  Data points are averaged over 100 random on-site energy realisations.}
 \label{3D_2}
\end{figure}

First of all, we re-consider the effect of random on-site energies $E_i$. Again, each point in Fig. \ref{enercN_2} and \ref{3D_2} corresponds to an average taken over 100 realisations. Figure \ref{enercN_2} shows that the loss in fidelity still scales linearly with the number of spins in the chain, but the effect observed is much more detrimental than it was for the simple two-qubit gate (see Fig. \ref{linear}) due to the increased number of excitations (now 4 vs. 2 in the two-qubit gate). Again, the decay in fidelity is not linear for increasing $\varepsilon/J_{max}$. While for very small values of $\varepsilon$, fidelity is very well maintained even for very long chains, a perturbation of as little as 5\% may already be very noticeable. Short chains up to $N=17$ achieve over 80\% of the fidelity, whereas longer chains such as $N=25$ suffer losses of over 30\%. Similarly, when $\varepsilon=0.1$ acceptable fidelity is only achievable for very short chains such as $N=9$, while chains of 17 spins and more become unsuitable for the proposed protocol as they achieve less than half of the desired fidelity.

In Fig. \ref{3D_2}, we see the combined influences of non-uniform on-site energies $E_i$, weighted by $\varepsilon$, and ${\cal{H}}'$ as given by Eq. (\ref{interc}) added to the Hamiltonian (\ref{hami}). Compared to Fig. \ref{decay}, the effect of $\varepsilon$ has become extremely detrimental, leading to a loss in fidelity of over 60\% for large values of $\varepsilon=0.2$. Even though a direct comparison of Fig. \ref{decay} and \ref{3D_2} is not possible due to the different measures of state transfer quality, it is safe to say that the crossed square cluster state is much more affected by perturbations due to unwanted non-uniform on-site energies. However, for smaller values of $\varepsilon$ up to a few percent, the fidelity of the knitted crossed square cluster is still very good at about 90\% of the ideal value. As the knitting protocol involves up to four excitations, the effect of ${\cal{H}}'$ is visible slightly more clearly but remains a very minor factor in the transfer quality, affecting the fidelity by a few percent only. The accuracy of Fig. \ref{decay} and \ref{3D_2} is the same, the perceived roughness of data points in Fig. \ref{decay}, which is due to the randomisation of the influence of $\varepsilon$, does also appear to the same extent in Fig. \ref{3D_2} but is less visible due to the different scale.

\begin{figure}
 \centering
 \includegraphics[width=0.48\textwidth]{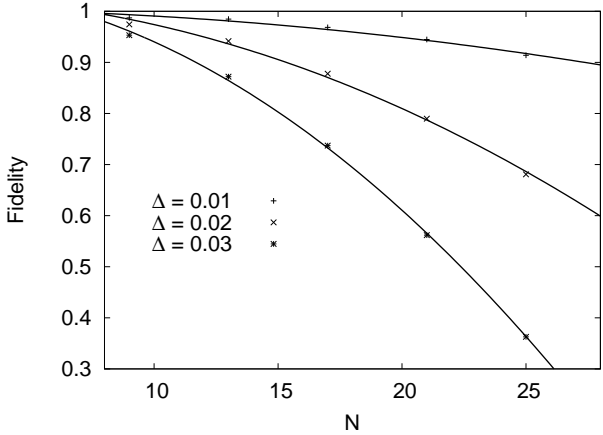}
 \caption{Fidelity of the ideal state at time $t=3t_{M}/2$ vs. $N$ for three values of $\Delta$, as labelled, where the chains are used to build a crossed square cluster state. Lines are quadratic fits to guide the eye.}
 \label{nnni2}
\end{figure}

Finally, we also re-consider the influence of next-nearest neighbour interaction, as given previously by Eq. (\ref{hami''}). Figure \ref{nnni2} shows that even for very small values of $\Delta$ below 5\%, there is a noticeable loss in state fidelity. The decay scales now as $N^2$ for the shown values of $\Delta$ and $N$, resulting in fidelities above 90\% only for $\Delta=0.01$ for chains of lengths up to 25 spins. For chains of 9 spins a next-nearest neighbour coupling $\Delta$ of up to 5\% still gives a fidelity of almost 90\% (not shown), while systems with 25 spins and $\Delta=0.03$ lose the vast majority of their fidelity. For larger values of $\Delta$, while the first fidelity peak at $t=3t_{M}/2$ might still be of acceptable value for short and medium length chains, it has to be noted that the periodicity is subsequently lost. Depending on the type of algorithm the obtained cluster state is intended for, this might pose a serious problem. For $\Delta=0.05$, chains of 21 spins or longer do not form fidelity peaks at $t=3t_{M}/2$ anymore, so that the protocol fails. This magnified detrimental effect of next-nearest neighbour interaction is again due to the increase in number of excitations, a phenomenon that can also be confirmed in non-entangling spin chains subject to this perturbation (not shown). This underlines the necessity for strict control of longer range interactions, particularly in long chains.

Another potential issue in the fabrication of distributed cluster ladders is that of untimely or non-synchronised injection of excitations into the spin chain. The effect of the delay of a single qubit on a two-qubit gate has been discussed in Ref \cite{me1}, but as the knitting of a distributed cluster state will involve four excitations most of the time, the possible delays are more involved. We will investigate the effect on the formation of a crossed square cluster state, the quality of which shall be measured via the state fidelity, and consider four separate delay scenarios. Referring to Fig. \ref{fig:squarecluster}, the system is subjected to a delay $\delta t$ as follows:

\begin{itemize}
 \item (A): Excitations 3 and 4 are both injected at $t_M/2+\delta t$
 \item (B): Excitation 4 is injected at $t_M/2+\delta t$ (while excitation 3 is on time)
 \item (C): Excitation 2 is injected at $\delta t$ and excitation 4 is injected at $t_M/2+\delta t$ (this corresponds to all injections on one side of the chain being delayed by the same amount)
 \item (D): Excitation 1 is injected at $\delta t$ and excitation 4 is injected at $t_M/2+\delta t$
\end{itemize}

\begin{figure}
 \centering
 \includegraphics[width=0.48\textwidth]{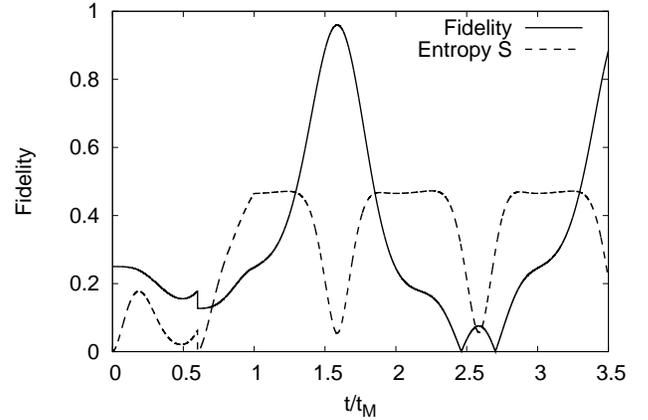}
 \caption{Effect of delay scenario A with $\delta t = 0.1t_M$ on the fidelity of a 9-spin chain.}
 \label{9ctd}
\end{figure}

Refocussing as discussed in section V is done immediately after each individual injection by measuring the auxiliary qubits. Despite an initial perturbation to the generation of the crossed square cluster state, we see in Fig. \ref{9ctd} (describing scenario A) that the evolution of the system continues to be essentially periodic if no read-out at $3t_M/2$ is undertaken. There is a clear kink in the plots of both entropy and state fidelity at $t_M/2+0.1t_M$, where $0.1t_M$ is the delay $\delta t$ of the injection of excitations 3 and 4. Consecutively, the fidelity peak occurs at $3t_M/2+\delta t$ and does not reach unity but a value of 0.9580, demonstrating the robustness of the system. 

\begin{figure}
 \centering
 \includegraphics[width=0.48\textwidth]{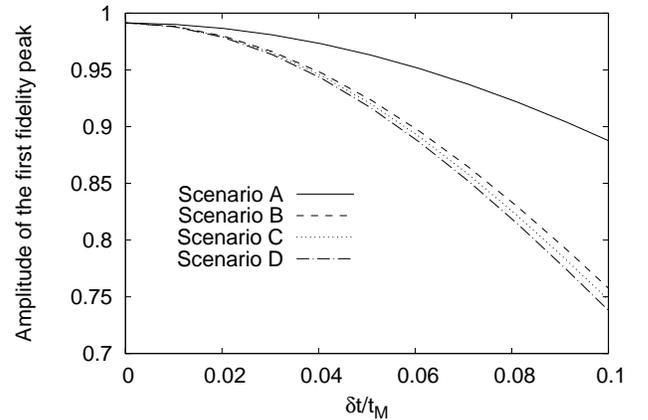}
 \caption{Effect of delay scenarios A-D on the amplitude of the first fidelity peak of a cluster state on a 9-spin chain.}
 \label{delay_comp}
\end{figure}

Figure \ref{delay_comp} shows the amplitude of the first fidelity peak at $3t_M/2+\delta t$ for a range of $\delta t$ and all four scenarios considered. First of all we note that none of the delay scenarios lead to fidelity of less than 90\% for a delay time of 5\% of the mirroring time $t_M$, while delay up to 10\% of $t_M$ might lead to a loss of over 25\% of fidelity. There is a clear discrepancy between scenario A and the other three scenarios, with scenario A performing much better and suffering just over 10\% fidelity loss for a delay of 10\% of $t_M$. Despite scenario B having fewer delayed excitations, it performs significantly worse. As such, it becomes clear that the system favours symmetrical input, also shown by scenario C with two excitations on the same end of the chain being delayed performing slightly better than scenario D, where excitations at opposite ends of the chain are delayed.

\section{Conclusions}

In conclusion, we have presented a method to knit distributed cluster states, using only a single spin chain set up for perfect state transfer and its natural dynamics. By closely observing the entropy of the chain, we have shown that the two end spins become decoupled from the rest of the chain at regular intervals, allowing us to inject further excitations into the chain without perturbing its existing excitation subspaces. The following natural chain dynamics lead the ensemble of excitations to entangle the end spins, which in turn decouple from the rest of the chain, allowing us to extract their states. As this routine of injections and extractions can be repeated without theoretical restrictions, we are thus producing a knitted cluster state consisting of an even arbitrary number of spins.

A further outlook of our technique is the possibility to knit other topological arrangements of cluster states by varying the injection and extraction timings and rates. As demonstrated in Fig. \ref{fig:knitting} and \ref{fig:squarecluster}, each crossing of excitations leads to a new edge in the final produced state. Provided that the chain used to knit an arbitrary arrangement is long enough to allow for the necessary localisation of excitations, the number of entanglement bonds produced is therefore freely controllable. Examples of topologically useful entangled states that could be achieved in this way are presented in reference \cite{lovett2011}. We would however expect more complicated structures which require very long chains, in particular those involving large numbers of excitations, to be more prone to the sources of decoherence we discussed.

We also examined multiple potential causes of error when knitting states, both for the fundamental building block of our protocol, the two-qubit gate, as well as for the smallest knitted cluster state, the crossed square cluster state. Consistent with existing results on the influence of errors on state transfer in spin chains, the two-qubit gate is very robust against unwanted interactions between different excitations as well as perturbations in the on-site energies, maintaining high levels of over 90\% of the desired entanglement even for large errors of 10\% and very long spin chains. Longer range interactions on the other hand have potentially very detrimental effects, with large perturbations leading to a significant decay in entanglement as well as a loss of periodicity. Consequently, we observed analogous phenomena for the crossed square cluster state, amplified greatly by the increased number of excitations. Here, only non-uniformity in the on-site energies of up to 5\% can be tolerated in order to ensure acceptable levels of over 70\% of formed entanglement in long chains. Again, it is however the next-nearest neighbour interaction that deserves most attention in the fabrication process of a spin chain, as even small unwanted interactions of just a few percent can lead to significant fidelity losses, as well as loss of periodicity of the system.

Additionally, we have considered the effect of the mistimed injection of one or more excitations on the formation of the desired crossed square cluster state and found that pairwise delay at the second injection time is much less detrimental than delay of a single excitation or delay of excitations at different injection times, but also that non-symmetric delay leads to a larger fidelity loss than delay restricted to the input on one end of the chain. Overall, the effect of delayed input as we considered is however quite limited and does not lead to a loss of more than 10\% of the perfect state fidelity for delays up to 5\% of $t_M$.

Our studies have analysed and demonstrated a new protocol for the fabrication of distributed cluster states which requires no additional resources or mechanisms beyond the set-up of a spin chain for perfect state transfer and the associated SWAP operations used to inject and extract information. In the context of limited errors deviating less than 10\% from the ideal set up, the knitting method proves to be a sound and stable method for generating distributed cluster states, making the utilisation of spin chains in quantum communication an ever more attractive prospect.

RR was supported by EPSRC-GB and Hewlett-Packard.


\begin{thebibliography}{99}
\bibitem{footnote} {Of course it must be appreciated that the chain length $N$ does enter the nearest neighbour coupling expression (\ref{PST}).}
\end{thebibliography}
\end{document}